\documentclass{article}
\usepackage{ltwol2e}

\def\NPB{{\em Nucl. Phys.} B}
\def\PLB{{\em Phys. Lett.}  B}
\def\PRL{{\em Phys. Rev. Lett.}}
\def\PRD{{\em Phys. Rev.} D}

\def\be{\begin{equation}}
\def\ee{\end{equation}}
\def\bea{\begin{eqnarray}}
\def\eea{\end{eqnarray}}

\begin{document}

\title{DUALITY IN STRING COSMOLOGY}

\author{RAM BRUSTEIN}

\address{Department of Physics, Ben-Gurion University, 
Beer-Sheva 84105, Israel\\E-mail: ramyb@bgumail.bgu.ac.il}   


\twocolumn[
\maketitle\abstracts{ Scale factor duality, a truncated form of time dependent
T-duality, is a symmetry of  string effective action in cosmological
backgrounds interchanging small and large scale
factors. The symmetry suggests a cosmological scenario (``pre-big-bang") in
which two duality related branches,  an inflationary branch and a decelerated
branch are smoothly joined into one non-singular cosmology.  
The use of scale factor duality
in the analysis of the higher derivative corrections to the effective action, and
consequences for the nature of exit transition, between  the inflationary and
decelerated branches, are outlined. 
A new duality symmetry is obeyed by the lowest order equations for 
inhomogeneity perturbations which always exist on top of the
homogeneous and isotropic background.    In some cases it  corresponds to a
time dependent version of S-duality, interchanging weak and strong coupling and
electric and  magnetic degrees of freedom, and in most cases  it corresponds to
a time dependent  mixture of both S-, and T-duality.
 The energy spectra obtained by using the new symmetry
 reproduce known  results of produced particle spectra, and can provide a useful
lower bound on particle production when our knowledge of the detailed dynamical
history of the background is approximate or incomplete.}
]

\section{Introduction}

Our starting point is the effective action (in the so-called ``string-frame"),
\begin{eqnarray}
S_{eff}&=&
\int d^4 x \left.\Biggl\{ \ \ \sqrt{-g} e^{-\phi}\left[ \frac{1}{16 \pi \alpha'}
\left(R+\partial_\mu\phi \partial^\mu\phi\right)\right.\right.  \nonumber \\
&-& \left. \left. \frac{1}{4} F_{\mu\nu}F^{\mu\nu}+ \bar\Psi D\hspace{-0.12in}\
 /\Psi
+\cdots \right] \right. \\ &+&\left. \hbox{\rm higher orders in $\alpha'
\partial^2$+ higher orders in $e^{\phi}$}    \right.\Biggr\}. \nonumber
\end{eqnarray}  
The dilaton $\phi$, is a ``Brans-Dicke-like" scalar with 
$\omega_{BD}=-1$.

The basic length scales of the theory are the string scale
$\alpha' \equiv  \ell_S^2$ and the Planck scale $ e^{\phi}\alpha' =  G_N\equiv
\ell_P^2$, related by
 $e^{\phi}\equiv g^2_{string}   \simeq \frac{1}{4 \pi} \alpha_{GUT}(1/\ell_S)=
\left(\frac{\ell_P}{\ell_S}\right)^2.
$  
 These relations are modified for strongly coupled string theory. We will
assume that the theory is weakly coupled throughout the evolution.

At early times fields may have been displaced  from  their present state, 
so we look for general FRW type solutions $g_{\mu\nu}=diag(-1, a^2(t) dx_i dx^i), 
\phi=\phi(t)$, and if other fields are present we assume that they have  only time
dependence.   We obtain equations  for the Hubble parameter $H(t)=\dot a(t)/a(t)$,
$\phi(t)$,  and additional fields, in particular moduli,   and matter (if
present), and solve them, requiring that
at late times the evolution has to  be that of standard cosmology. 

In a general effective action we may represent any contributions in  addition to
lowest order action by a ``matter" Lagrangian \cite{sfd,pbb,bm1}
\begin{eqnarray}
S_{eff}&=&
\int d^4 x \left\{ \ \ \sqrt{-g}\left[ \frac{e^{-\phi}}{16 \pi \alpha'}
\left(R+\partial_\mu\phi \partial^\mu\phi\right)\right]\right.  \nonumber \\
&+& \left.{\frac{1}{2}} {\cal L}_m(\phi,g_{\mu\nu},...)   \right\}. 
\end{eqnarray}  
The equations of motion are the following
\begin{eqnarray}
&&\dot \phi = 3H_S\pm\sqrt{3H_S^2+e^\phi \rho_S} \nonumber\\
&&\dot H_S=\pm H_S\sqrt{3H_S^2+e^\phi \rho_S}+
{\frac{1}{2}} e^\phi( p_S+\Delta_\phi{\cal L}_m) \\
&&\dot\rho_S+3 H_S(\rho_S+p_S)= -\Delta_\phi{\cal L}_m \dot\phi, \nonumber
\label{fstordm}
\end{eqnarray}
where
\begin{eqnarray}
&&T_{\mu\nu} = \frac{1}{\sqrt{-g}} \frac{\delta {\cal L}_m}{\delta g^{\mu\nu}}  \nonumber\\
&& \Delta_\phi{\cal L}_m={\frac{1}{2}} \frac{1}{\sqrt{-g}}  \frac{\delta {\cal L}_m}{\delta
\phi} \\  &&T^{\mu}{}_{\nu}=diag(\rho,-p,p,-p).\nonumber
\end{eqnarray}

As a result of scale factor duality (SFD) which will be discussed in more detail below,
the solutions come in pairs or branches, the $(+)$ branch vacuum (without any sources)
satisfies $$\dot H_S=+H_S\sqrt{3H_S^2},$$
$H_S=\frac{1}{\sqrt{3}} \frac{1}{t_0-t},\ 
 t<t_0$ and is characterized by a future singularity. If the universe  starts
expanding according to the $(+)$ branch vacuum solution, $H$ is positive, and
therefore its time derivative is also positive. This branch cannot connect smoothly to
radiation dominated  (RD) FRW with constant dilaton,   $\dot \phi = 3H_S+\sqrt{3H_S^2+e^\phi \rho_S}
\Rightarrow\dot \phi>0 $. The $(-)$ branch vacuum satisfies 
$$\dot H_S=-H_S\sqrt{3H_S^2},$$
$ H_S=\frac{1}{\sqrt{3}},\ \frac{1}{t-t_0} t>t_0$
and is characterized by a past singularity.
This branch can connect smoothly to RD FRW with constant
dilaton  $$\dot \phi = 3H_S-\sqrt{3H_S^2+e^\phi \rho_S}
\Rightarrow\dot \phi=0\ \hbox{\rm if}\ 6H_S^2=e^\phi \rho_S.$$

A cosmological scenario (``pre-big-bang") in which two duality related
branches,  an inflationary branch and a decelerated branch are smoothly joined
into one cosmology has been proposed \cite{sfd,pbb}.   In this scenario the universe
quickly becomes homogeneous, isotropic,  and spatially  flat. 
The transition between the inflationary and decelerated branches, the
so-called graceful exit transition, is expected to occur when the universe
reaches string scale curvature. The use of scale factor duality in the analysis
of the higher derivative corrections to the effective action, and consequences
for the nature of exit transition are outlined.

Quantum fluctuations superimpose on top of the
smooth classical background  inhomogeneity perturbations, which are then
amplified by the accelerated expansion of  the universe and materialize as
particles with specific energy spectra later on \cite{mukh}. 

A new duality symmetry is obeyed by the lowest order equations for 
inhomogeneity perturbations \cite{sduality}.  In some cases it 
corresponds to a time dependent version of T-duality, interchanging small and
large scale factors, in some cases it  corresponds to a time dependent version of
S-duality, interchanging weak and strong coupling and electric and 
magnetic degrees of freedom, and in most cases  it corresponds to a
time dependent  mixture of both. As in other applications, duality turns out to
be a powerful tool  for obtaining results that are inaccessible otherwise. 
In particular, lower bounds on the energy density of the produced particles.
The energy spectrum obtained by truncating the
solutions of the perturbation equations to the constant modes, 
is characterized by a residual duality symmetry, reproduces known  results of
produced particle spectra, and can provide a useful lower bound on particle
production when our knowledge of the detailed dynamical history of
the background is approximate or incomplete.

\section{Scale Factor Duality}
\subsection{Lowest order}\label{subsec:lstord}

To introduce SFD we look at the lowest order 4-d effective  dilaton-gravity action
\begin{eqnarray}
 S_{LO}&=&
\int d^4 x \ \ \sqrt{-g}\left[ \frac{e^{-\phi}}{16 \pi \alpha'}
\left(R+\partial_\mu\phi \partial^\mu\phi\right)\right].  \\
 &&\hspace{-.5in} \hbox{\rm Integration by parts leads to} \nonumber \\
S_{LO}&=& -\frac{1}{16 \pi \alpha'}\int d^4 x a^3 e^{-\phi}
\left(6 H^2 -6 H \dot \phi + \dot \phi^2\right) \nonumber \\
&=& -\frac{1}{16 \pi \alpha'}\int d^4 x e^{-\bar{\phi}} \left(3 H^2- 
\dot {\bar{\phi}}^2\right),
\label{slo}
\end{eqnarray} 
where $\bar{\phi}\equiv \phi -3 \ln a $, and 
we have set the lapse function $n(t)$ to unity.

The action (\ref{slo}) is invariant under the symmetry transformation
\begin{eqnarray}
a(t) &  \rightarrow & 1/a(t), \nonumber \\
\phi(t) & \rightarrow & \phi-6 \ln a(t),
\nonumber
\end{eqnarray} 
$\bar\phi$ and $H^2$ are invariant, 
$$
 {\bar \phi(t)} \rightarrow  {\bar \phi(t)},$$
$$
 {H(t) \rightarrow -H(t),}
$$
and the equations of motion are covariant. The two branches describing an expanding
universe are related to each other SFD$\times$ Time reversal. 
In general, for more complicated cosmological backgrounds the symmetry of the action 
is more complicated \cite{odd}.

\subsection{Leading corrections}\label{subsec:2ndord}

The transition between the two duality related branches is called the graceful exit
transition. It is known that to lowest order the two branches are separated by a
singularity \cite{exit}. The emerging scenario for the exit transition requires classical
$\alpha'$ corrections  which can bound the curvature
below the string scale, as well as quantum corrections \cite{gmv,bm2}. 
The leading classical corrections determine whether the
solution can reach a ``good" region in  $\dot {\bar{\phi}}, H$ phase space. A model
for the exit transition has been presented \cite{bm2}, and therefore we know that a
transition is possible. The question is whether string theory actually determines
the coefficients such that a transition occurs. 

We have investigated \cite{bm3} effective
classical corrections, and demanded that the action will really be an
effective dilaton-gravity action,  without additional  new degrees 
of freedom \cite{fotw}. This
means that we have to use actions that produce equations without higher derivatives.
Field redefinitions can change that but, it is better to use a ``frame" with no
higher derivatives ensuring numerical stability and control. We
also require an action that  reproduces whatever string theoretic information available
such as  scattering amplitudes, perturbative beta-function calculations etc. 
As we will see this is not enough to obtain the full corrected action. If scale factor duality
could be imposed in a practical way it would help, however, the situation is more 
complicated.

The action including leading classical corrections is given by
$$
S_{LCC}= \int {d^4 x} \sqrt{g} {e^{-{\phi}}} \{
 R
 + {{(\nabla \phi )}^2}
 +$$
$$
 {1 \over 2} [
 {A\,{(\nabla \phi )^4}}
 +
 {B\,{R^2}_{GB}}
 + $$
$$
 {C\,(R^{\mu
 \nu}-{1 \over 2} g^{\mu \nu} R) \,\nabla_{\mu} \phi \,\nabla_{\nu} \phi}
 + 
 {D\,\nabla^2(\phi)\,{{(\nabla \phi)}^2}}
 ] \}.
$$
In covariant variables, it takes the following form
\begin{eqnarray}
S_{LCC}={e^{-{\bar \phi}}} &&
 \biggl\{ {{3\,{{H(t)}^2}}\over {n(t)}} 
 -
 {{{{{\dot 
 {\bar  \phi}}(t)}^2}}\over {n(t)}} 
\nonumber \\
&&
 +
 {{3\,\left(
 27\,A 
 + 
 8\,B 
 + 
 9\,{
 C} 
 + 
 27\,{
 D} 
 \right) 
 \,{{H(t)}^4}}\over 
 {2\,{{n(t)}^3}}} 
  +\nonumber \\
&&
 {{\left(
 54\,A 
 + 
 4\,B 
 + 
 9\,{
 C} 
 + 
 45\,{
 D} 
 \right) 
 \,{{H(t)}^3}\,{\dot 
 {\bar 
 \phi}}(t)}\over 
 {{{n(t)}^3}}} 
\nonumber \\
&&
 + 
 {{3\,\left( 
 18\,A 
 + 
 {
 C} 
 + 
 12\,{
 D} 
 \right) 
 \,{{H(t)}^2}\,{{{\dot 
 {\bar 
 \phi}}(t)}^2}}\over 
 {2\,{{n(t)}^3}}} 
 + \nonumber \\
&&
 {{3\,\left( 
 2\,A 
 + 
 {
 D} 
 \right) 
 \,H(t)\,{{{\dot 
 {\bar 
 \phi}}(t)}^3}}\over 
 {{{n(t)}^3}}} 
\nonumber \\
&&
 + 
 {{\left( 
 3\,A 
 + 
 {
 D} 
 \right) 
 \,{{{\dot 
 {\bar 
 \phi}}(t)}^4}}\over 
 {6\,{{n(t)}^3}}} 
 \biggr\}, \nonumber
\end{eqnarray}
where we have performed integration by
parts to get rid of the $h'(t)$, $\phi''(t)$ and $n'(t)$. This is possible
in general due to the 'no higher derivatives' condition.

Perturbative string calculation can provide two of the coefficients, 
one additional coefficient sets the
overall scale at which the leading corrections kick in, so one coefficient 
remains unknown. If some symmetry principle, such as SFD could be used, 
the leading corrections action would be determined completely.
For example, to impose naive SFD we would have to eliminate the 
odd parts of the action. We can set, for example,
$$
D=-2 A $$ $$
C=4 A-{4 \over 9} B.
$$

However, this is only possible in a homogeneous background \cite{hd}.
In an inhomogeneous background we get more equations and the only consistent
solution is $A=B=C=D=0$. The conclusion is that, at the very least, 
 the SFD transformation  has to be modified at this order, making it less useful for 
determination of the one remaining coefficient.
If we insist on naive SFD in the homogeneous case, then it is possible to show 
\cite{bm3} that if a stable algebraic fixed point exists, another non-stable 
fixed point will also exist, and that the generic solution will encounter the 
unstable fixed  point first and run into a singularity. 

\subsection{All orders}\label{subsec:allord}

As we have seen, additional input is required to determine the behaviour of solutions
when classical stringy corrections are included. The best would be to establish
the existence of an exact conformal field theory solution corresponding to the 
algebraic fixed point. In general this requires
working with 2-d conformal field theories rather than with effective actions.
I outline here some of the possibilities to achieve progress in that 
direction \cite{bm3}.

For highly symmetric backgrounds, such as the linear-dilaton deSitter
background, it is possible to use the isometries of the background to impose
additional symmetries on world-sheet operators, and constrain the beta-function 
coefficients.
Another possibility is to use conformal perturbation theory to add (1,1) operators to
an established conformal field theory, a linear dilaton flat-space 
background \cite{bm3}. 
Yet another possibility is to start with known exact solutions in higher
dimensions and compactify down to 4-d \cite{ck}.

\section{``S"-Duality}

The quadratic action for perturbations of any tensor field expanded around a
cosmological dilaton-gravity background is given by 
\begin{equation}
S_{pert}=\frac{1}{2} \int d^3 x d\eta S(\eta) \left[\psi'^2-(\nabla\psi)^2\right].
\label{spertact}
\end{equation}
The prefactor $S(\eta)$ is given by $a^{2m} e^{\ell \phi}$, where $m$ and $\ell$
depend on the type of field. For example, gravitons, dilatons and moduli have $m=1$,
$\ell=-1$, model independent axions have $m=1$, $\ell=1$, while Ramond-Ramond axions
have $m=1$, $\ell=0$, and so on. We would like to compute the evolution of
perturbation and eventually compute an important physical observable quantity: the
spectrum of produced particles at late times.

\subsection{``S"-Duality symmetry}

To discuss duality symmetry of the action (\ref{spertact}) it is more convenient to
to use the Hamiltonian formalism.
The Hamiltonian density corresponding to (\ref{spertact}) is given by
\begin{equation}
 H=\frac{1}{2} \int d\eta \Biggl\{  S^{-1} \Pi^2 + 
S (\nabla\Psi)^2\Biggr\},
\label{ham}
\end{equation}
where the momentum conjugate to $\Psi$ is given by
\begin{equation}
 \Pi= S \Psi'.
\label{pipsi}
\end{equation}
The Hamilton equations of motion are first order
\begin{eqnarray}
\Pi'&=&-\frac{\delta H}{\delta \Psi}= S \nabla^2\Psi \nonumber \\
\Psi'&=&\frac{\delta H}{\delta \Pi}=  S^{-1} \Pi, 
\label{frstord}
\end{eqnarray}
and lead to second order equations
\begin{eqnarray}
\Pi''-\frac{S'}{S} \Pi'- \nabla^2\Pi&=&0 \label{pieq} \\
\Psi''+\frac{S'}{S} \Psi'- \nabla^2\Psi&=&0. \label{psieq} 
\end{eqnarray}
The second equation (\ref{psieq}) is commonly used in analysis of
perturbation spectra.

In Fourier space the Hamiltonian density is given by 
\begin{equation}
H=\frac{1}{2} \int d\eta \Biggl\{  S^{-1} \Pi_{\vec{k}}\Pi_{-\vec{k}} + 
S k^2 \Psi_{\vec{k}}\Psi_{-\vec{k}}\Biggr\},
\label{fham}
\end{equation}
and the equations of motion are given by 
\begin{eqnarray}
\Pi_{\vec{k}}'&=&-S k^2 \Psi_{-\vec{k}} \nonumber \\
\Psi_{\vec{k}}' &=&S^{-1}{\Pi}_{-\vec{k}}. 
\label{ffrstord}
\end{eqnarray}

``S"-duality exchanges the variables and momenta and at the same time sends $S$ to
its inverse,
\begin{eqnarray}
\Pi_{\vec{k}}&\rightarrow&\widetilde \Pi_{\vec{k}}=k \Psi_{\vec{k}}
 \nonumber \\
k \Psi_{\vec{k}} &\rightarrow&k \widetilde{\Psi}_{\vec{k}}=-\Pi_{\vec{k}} 
\nonumber \\
S &\rightarrow&\widetilde S=S^{-1},
\label{sduality}
\end{eqnarray}
leaving the Hamiltonian, equations of motion and Poisson brackets invariant. 

We are interested in a situation in which the initial conditions correspond to 
zero-point vacuum fluctuations of the field $\Psi$, and therefore 
\begin{equation}
\langle S^{-1} \Pi^2\rangle= \langle S (\nabla\Psi)^2\rangle,
\label{incon}
\end{equation}
where $\langle\cdots\rangle$ denotes ensemble average.

 The duality (\ref{sduality}), contains strong-weak coupling duality
as a special case.  For perturbative heterotic 4-d gauge bosons the function $S$
is given simply by $S(\eta)=e^{\phi(\eta)}$. Recall that
$e^{\phi(\eta)}=g_{string}$, so the transformation   $S \rightarrow \widetilde
S=S^{-1}$ is, at each time $\eta$, simply the celebrated strong-weak coupling
duality $g_{string}\rightarrow g_{string}^{-1}$, which appears as a part of the
$SL(2,Z)$ group, usually called $S$-duality. The transformation (\ref{sduality})
exchanges in this case electric and magnetic degrees of freedom.

\subsection {Approximate solutions}

To construct approximate solutions define ${\widehat\Psi}$, ${\widehat\Pi}$,
whose Fourier modes are given by
\begin{eqnarray}
\widehat\Psi_k &=&{S}^{1/2}\ \Psi_k \nonumber \\
\widehat\Pi_k &=&{S}^{-1/2}\ \Pi_k. 
\label{nv}
\end{eqnarray}
The new variables have simple transformation law under ``S"-duality
\begin{eqnarray}
k\widehat\Psi_k&\rightarrow&k\widetilde{\widehat{\Psi}}_k=-\Pi_k \nonumber \\
\widehat\Pi_k &\rightarrow&\widetilde{{\widehat\Pi}}_k=k\Psi_k \nonumber \\
S &\rightarrow&\widetilde S=S^{-1}.
\label{ksduality}
\end{eqnarray}

The variables ${\widehat\Psi}$, ${\widehat\Pi}$
 satisfy the following Schr\"odinger-like equations
\begin{eqnarray}
{\widehat\Psi}_k{''}+\left(k^2-(S^{1/2}){''} S^{-1/2}\right){\widehat\Psi}_k 
&=&0 \nonumber \\
{\widehat\Pi}_k{''}+ \left(k^2-(S^{-1/2}){''} S^{1/2}\right)
{\widehat\Pi}_k&=&0. 
\label{nscndord}
\end{eqnarray}
Since $S(\eta)\sim\eta^\alpha$, the potentials $V_{\Psi}=(S^{1/2}){''}
S^{-1/2}$, $V_{\Pi}=(S^{-1/2}){''} S^{1/2}$, if non-vanishing, 
 are proportional to $1/\eta^2$.
For $k^2> V_{\Psi}$, $V_{\Pi}$, or equivalently $(k\eta)^2>1$
(inside the horizon),  we look for WKB-like approximate solutions
\begin{eqnarray}
{\widehat\Psi}_k(\eta)&=&\left( k^2-V_{\Psi}\right)^{-1/4}\ 
e^{\ -i\int\limits_{\eta_0}^{\eta} d\eta' \left( k^2-V_{\Psi}\right)^{1/2} }
 \nonumber \\
{\widehat\Pi}_k(\eta)&=& k \left( k^2-V_{\Pi}\right)^{-1/4}\ 
e^{\ -i\int\limits_{\eta_0}^{\eta} d\eta' \left( k^2-V_{\Pi}\right)^{1/2} }.
\label{wkbsol}
\end{eqnarray}
 The advantage of looking at
solutions (\ref{wkbsol}) is that they manifestly  preserve the ``S"-duality
symmetry of the equations, because the potentials  $V_{\Psi}$, $V_{\Pi}$ get
interchanged under $S \rightarrow S^{-1}$. 

For very large $k^2$, $k^2\gg V_{\Psi}$, $V_{\Pi}$ 
solutions (\ref{wkbsol}) reduce to correctly normalized vacuum fluctuations
\begin{eqnarray}
{\widehat\Psi}_k(\eta)&=& k^{-1/2}\ 
e^{\ -i k\eta+ i\varphi_0 } \nonumber \\
{\widehat\Pi}_k(\eta)&=& k^{+1/2}\ 
e^{\ -i k\eta+ i\varphi'_0 },
\label{wkbsol1}
\end{eqnarray}
where $\varphi_0$, $\varphi_0'$ are random phases, originating from the random
initial conditions. Note that because of the random phases, ``S"-duality holds only
on the average in the sense of eq.(\ref{incon}).

For $k^2< V_{\Psi}$, $V_{\Pi}$, or equivalently $(k\eta)^2<1$ (outside the
horizon), it is possible to write ``exact" solutions \cite{sduality}.
It is convenient  to define the functions
$T\!\cos(S^{-1},S)$,  $T\!\sin(S^{-1},S)$ 
\begin{eqnarray}
&&\hspace{-.2in}T\!\cos(S^{-1},S)=
1- k \int\limits_{\eta_{ex}}^\eta\!\!\! d\eta_1 S^{-1}(\eta_1) 
\ k \int\limits_{\eta_{ex}}^{\eta_1}\!\!\! d\eta_2 S (\eta_2) 
+\cdots \nonumber \\
&+& 
(-1)^{n+1} k^{2(n-1)}\hspace{-.1in} \prod_{n-1\  times}\hspace{-.2in}
 \int\!\!  S^{-1} \int\!\! S\cdots 
\int\!\!  S^{-1} \int\!\! S
+\cdots\nonumber \\
&&\hspace{-.2in}T\!\sin(S^{-1},S)=
k \int\limits_{\eta_{ex}}^\eta d\eta_1 S^{-1}(\eta_1)
- \cdots  \\
&+& 
(-1)^{n+1} k^{2n-3}\!\!\int  S^{-1} \hspace{-.2in}
\prod_{n-2\  times}\hspace{-.2in}
  \int\!\! S\!\! \int\!\!  S^{-1}\!\! \cdots\int\!\! S \!\!\int\!\!  S^{-1}\!\!
+\cdots, \nonumber
\label{tsincos3}
\end{eqnarray}
in terms of  which the ``exact" solutions take the following form
\begin{eqnarray}
{\widehat\Psi}_k(\eta)&=& \sqrt{S}\Biggl\{ 
A_k\ T\!\cos(S^{-1},S)\!\!+\!\!  B_k\ T\!\sin(S^{-1}\!\!,S)\Biggr\}\nonumber \\
{\widehat\Pi}_k(\eta)&=& \frac{k}{\sqrt{S}} \Biggl\{ 
B_k\ T\!\cos(S,S^{-1})\!\!-\!\!A_k\ T\!\sin(S,S^{-1})  \Biggr\}.\ \ \ \ \ \
\label{wkbsol3}
\end{eqnarray}
Using the relations
\begin{eqnarray}
 \left[T\!\cos(S^{-1},S)\right]'&=& -\frac{k}{S} \ T\!\sin(S,S^{-1})\nonumber \\
 \left[T\!\sin(S^{-1},S)\right]'&=& \frac{k}{S} \ T\!\cos(S,S^{-1}),
\label{tsc}
\end{eqnarray}
and similar relations for $\left[T\!\cos(S,S^{-1})\right]'$ and 
$\left[T\!\sin(S,S^{-1})\right]'$, it is possible to verify explicitly that
${\widehat\Psi}_k$,  ${\widehat\Pi}_k$ in eq.(\ref{wkbsol3}) are indeed
solutions of eqs.(\ref{nscndord}). Formally, these solutions are valid also
inside the horizon,  but the functions $T\!\cos$, $T\!\sin$ are not
well defined there.

We need to match the solutions inside and outside the horizon and  do it such
that ``S"-duality is respected. One way of doing so is to use  solutions 
(\ref{wkbsol}) inside the horizon, and (\ref{wkbsol3}) outside the
horizon and match them at some time near
horizon exit time $\eta_{ex}$, for which $k\eta_{ex}\sim 1$. Taking 
advantage of the phenomenon of ``freezing of perturbations" outside 
the horizon we obtain the following result,
\begin{eqnarray}
  {\widehat\Psi}_k(\eta)  &=&
\frac{1}{\sqrt{k}} \left[\left({S_{ex}\over S_{re}}\right)^{\!-1/2}\hspace{-.25in}
\cos (k\eta)  +  \left({S_{ex}\over S_{re}}\right)^{1/2}\hspace{-.15in}
\sin(k\eta)  \right] 
\nonumber \\
{\widehat\Pi}_k(\eta) &=&
\sqrt{k} \left[ \left({S_{ex}\over S_{re}}\right)^{1/2}\hspace{-.2in}
\cos (k\eta) - \left({S_{ex}\over S_{re}}\right)^{\!-1/2}\hspace{-.2in}
\sin(k\eta)\right] 
\label{appsol}
\end{eqnarray}
where $S_{re}=S(\eta_{re})$. The reentry time $\eta_{re}$ is the second time at which
$k\eta_{re}\sim 1$.

\subsection{Energy Spectrum}

We compute an important physical observable, the Hamiltonian density,
\begin{equation}
\langle H_k\rangle ={1\over 2}\left(\langle|{\widehat\Pi}_k|^2\rangle
+k^2 \langle  |{\widehat\Psi}_k|^2\rangle \right)
\end{equation}
Using the approximate solutions (\ref{appsol}) we obtain

\begin{equation}
\langle H_k\rangle = k\left(
{S_{ex}\over S_{re}} + {S_{re}\over S_{ex}}\right).
\end{equation}
It is invariant  under  $S_{ex}\rightarrow S_{ex}^{-1}$, $S_{re}
\rightarrow S_{re}^{-1}$, and  overall rescaling of $S$. Note that, for a given $k$,
$\langle H_k\rangle$ depends only on $S_{ex}$ and $S_{re}$, and not on the 
whole evolution.

The spectral energy distribution,
$d\rho_k/d\ln k$ = $(k^3/a^4) \langle H_k\rangle$  ($\omega =k/a$) 
is given by
\begin{eqnarray}
 {d\rho(\omega)\over d  \ln \omega} &=& \omega^4 \left[
{S_{ex}(\omega)\over S_{re}(\omega)} + {S_{re}(\omega)
\over S_{ex}(\omega)}\right]\nonumber \\
&\simeq&  \omega^4 ~ {\rm Max}
\Biggl\{ {S_{ex}\over S_{re}} , {S_{re}\over S_{ex}}\Biggr\}.
\label{apprho}
\end{eqnarray}
It has the same invariance properties as the Hamiltonian density.

From eq.(\ref{apprho}) we obtain model independent lower bound  on 
 energy density of cosmologically produced particles. The spectrum (\ref{apprho})
is a sum of two terms, one being the inverse of the other. Therefore it is not 
possible to decrease the contribution of one term without increasing the contribution of
the other. The physical origin of this lower bound is indeed the uncertainty principle. 
Recall that one term originates from the contribution of the perturbation conjugate 
momentum and the
other from the contribution of the
perturbation itself. The uncertainty principle says that it is not possible 
to decrease both without limits.
For specific cases, the lower bound may be improved using some particular properties
of the background.

The result (\ref{apprho}) provides an easy and a very general way of computing 
${d\rho(\omega)\over d  \ln \omega}$. The prescription is simple. 
Once the function $S(\eta)$ is known for all times, substitute 
for $\eta_{ex}\rightarrow k^{-1}$, and for $\eta_{re}$ substitute the properly 
redshifted $k^{-1}$. The results obtained using this simple method reproduce
known results obtained by explicit complicated calculations. 
\cite{dilatons,gravitons,photons,axions,gnrl,lrgscle}

\section*{Acknowledgements}
I would like to thank my collaborators Maurizio Gasperini, Merav Hadad,  Dick
Madden and Gabriele Veneziano. This work is supported in part by the  
Israel Science Foundation administered by the Israel Academy of Sciences 
and Humanities.

\end{document}